\documentclass[aps,pr,twocolumn,noshowpacs,superscriptaddress,longbibliography]{revtex4-2}
\usepackage{amsmath,graphicx,amsbsy,amssymb,amsmath,epsfig,latexsym,wasysym,pifont,mathrsfs,natbib}
\usepackage{float} 
\newfloat{widefig}{thp}{lop} 
\usepackage{comment}
\usepackage{soul}
\usepackage{array, makecell} \usepackage{verbatim} \usepackage{datetime}
\usepackage{multirow,array} \usepackage{hyperref} \usepackage{color} \usepackage{setspace}
\usepackage{graphicx}
\usepackage{graphics}
\usepackage{xcolor} 
\usepackage{dirtytalk}
\usepackage[british]{babel}
\usepackage[utf8,latin1]{inputenc}
\UseRawInputEncoding
\usepackage{csquotes}
\usepackage{comment}
\usepackage[export]{adjustbox}
\usepackage{siunitx}
\usepackage[version=4]{mhchem}
\usepackage{amsmath}
\usepackage{amssymb}
\usepackage{mathtools}

\hypersetup{colorlinks,
  linkcolor=blue,%
  citecolor=blue,%
  urlcolor=blue}

\begin{document}


\title{\textbf{Revisiting thermal transport in \ce{CuCl}: First-principles calculations and machine learning force fields}}






\author{Ashis Kundu}
\email[]{ashis.kundu@liu.se}
\affiliation{Theoretical Physics Division, Department of Physics, Chemistry and Biology (IFM), Linköping University, SE-581 83 Linköping, Sweden}
\author{Florian Knoop}
\affiliation{Theoretical Physics Division, Department of Physics, Chemistry and Biology (IFM), Linköping University, SE-581 83 Linköping, Sweden}
\author{Igor A. Abrikosov}
\affiliation{Theoretical Physics Division, Department of Physics, Chemistry and Biology (IFM), Linköping University, SE-581 83 Linköping, Sweden}

\date{\today}


\begin{abstract}  

Accurate prediction of lattice thermal conductivity ($\kappa_l$) in strongly anharmonic materials requires renormalized interatomic force constants (IFCs) and appropriate incorporation of diagonal and off-diagonal contributions and higher-order scattering. We investigate \ce{CuCl}, a highly anharmonic system with a simple zincblende structure and ultralow $\kappa_l$. Our calculations,  including IFC renormalization and four-phonon scattering, show excellent agreement with the experiment, underscoring the critical role of both effects in the accurate estimation of $\kappa_l$. Furthermore, the unusual pressure dependence of $\kappa_l$ is explored using a rigorously validated machine-learned force field, with the predicted values showing good agreement with the experimentally observed trend of monotonic decrease. This behavior is primarily driven by a significant increase in four-phonon scattering and a reduction in the group velocity of transverse acoustic modes. Overall, this study establishes a robust framework for modeling thermal transport in strongly anharmonic materials.

\end{abstract}


\maketitle

\section{Introduction}

Understanding and an accurate evaluation of lattice thermal conductivity in highly anharmonic materials is essential for thermoelectric~\cite{Snyder_NM08,Sootsman_ACIE09} and thermal barrier coatings~\cite{Nitin_SC02,Clarke_MRS12}, as these materials typically have low thermal conductivity. Binary semiconductors generally tend to have high thermal conductivity due to their simple crystal structures and phonon characteristics, which stem from large atomic mass differences that favour weak phonon scattering~\cite{Feng_PRB16, Feng_PRB17, Ravichandran_PRX20,Xia_PRX20}. However, some compounds exhibit strong anharmonicity despite their simple structures and significant mass difference, leading to ultralow thermal conductivity~\cite{Xia_PRX20}. One such example is \ce{CuCl}, a metal halide with a zincblende crystal structure, which has been experimentally reported to exhibit a thermal conductivity of approximately 0.7 to 0.85 W m$^{-1}$ K$^{-1}$ at room temperature~\cite{Slack_PRB82,Haynes_CRC14}.
Previous theoretical studies have attempted to explain this unusually low value and attributed it to extreme anharmonicity caused by strong orbital hybridization~\cite{Togo_PRB15,Mukhopadhyay_PRB17,Yang_SCPMA23}. However, the calculated thermal conductivity varies, with some studies overestimating~\cite{Togo_PRB15,Mukhopadhyay_PRB17} while others underestimating~\cite{Yang_SCPMA23} the experimental result. This discrepancy highlights the need for a more accurate assessment of anharmonicity and the lattice thermal conductivity.

Another peculiar aspect of the thermal conductivity of \ce{CuCl} is its pressure dependence. Typically, thermal conductivity increases with pressure due to phonon hardening unless a structural phase transition occurs~\cite{Zhou_NRP22}. However, recent studies have reported anomalous behavior in some materials, where thermal conductivity initially increases and then decreases with pressure~\cite{Lindsay_PRB15,Ouyang_15,Yuan_PRB18,Ravichandran_NC19,Ravichandran_NC21,Kundu_PRL24}. This behavior results from the interplay between harmonic and anharmonic interactions or competition among different anharmonic scattering mechanisms. In the case of \ce{CuCl}, thermal conductivity continuously decreases with pressure before the phase transition~\cite{Slack_PRB82}. This open question requires further investigation.

In this work, we study the lattice thermal conductivity of \ce{CuCl} theoretically using Temperature Dependent Effective Potential (TDEP) method~\cite{Hellman_PRB11,Hellman_PRB13-2} accurately incorporating anharmonic effects, which results in excellent agreement with the experiment. Our findings highlight the importance of both temperature-dependence of interatomic force constants (IFCs) and four-phonon interactions. Moreover, we study 
the significant role of selection of the exchange-correlation functional in the density functional theory (DFT) calculations of \ce{CuCl}. To study pressure-dependent thermal conductivity, we employ a machine-learning force field (MLFF) to reduce computational costs while ensuring reliability. The accuracy of MLFF is verified at multiple levels, and our calculated thermal conductivity across temperatures and pressures matches well with experimental results. Thus, our work provides a robust framework for estimating thermal conductivity in highly anharmonic systems.


\section{Methodology}

To compute the lattice thermal conductivity tensor, we employ the mode-coupling theory of anharmonic lattice dynamics~\cite{Castellano_JCP23,Batista_PRB25}, which provides a nonperturbative framework that incorporates phonon correlation functions and captures both collective and coherent contributions to thermal transport, as implemented in the TDEP package~\cite{TDEP-package}. The total thermal conductivity ($\kappa$) is the sum of the diagonal and off-diagonal (coherent) contributions, derived from the respective heat flux operators, and is expressed as
\begin{equation}
    \kappa = \kappa^{\text{d}} + \kappa^{\text{nd}}.
\end{equation}

The diagonal contribution to the thermal conductivity tensor is expressed as
\begin{equation}
\kappa_{\alpha\beta}^{\mathrm{d}} = \frac{1}{V} \sum_{\lambda\lambda'} v_{\lambda}^{\alpha} v_{\lambda}^{\beta} c_\lambda \mathbf{\Sigma}^{-1}(\lambda,\lambda'),
\end{equation}
where \(V\) is the volume of the primitive cell, and \(\alpha\) and \(\beta\) are Cartesian axes. The phonon mode \(\lambda \equiv (\mathbf{q}, s)\) is labeled by the phonon wave vector \(\mathbf{q}\) and the branch index \(s\). \(v_{\lambda}^{\alpha}\) is the group velocity of mode \(\lambda\), and \(c_{\lambda} = n_{\lambda} (n_{\lambda} + 1) \omega_{\lambda}^{2} / k_{\mathrm{B}} T^2\) is the heat capacity, where \(\omega_{\lambda}\) is the phonon frequency, \(n_{\lambda}\) is the Bose-Einstein occupation number, \(k_{\mathrm{B}}\) is the Boltzmann constant, and \(T\) is the temperature. The matrix \(\mathbf{\Sigma}\) denotes the phonon scattering matrix, whose diagonal elements correspond to the phonon scattering rates \(\Gamma_{\lambda}\), that incorporates contributions from three-phonon, four-phonon, and phonon-isotope scattering. The off-diagonal elements of \(\mathbf{\Sigma}\)  describe coupling between modes that gives collective phonon contributions to heat transport. The calculation of \(\kappa_{\alpha\beta}^{\mathrm{d}}\) in this form is computationally demanding, as it requires the diagonalization of the scattering matrix. This difficulty is alleviated using an iterative solution approach, as detailed in Ref.~\onlinecite{Batista_PRB25}.

The off-diagonal (coherent) contribution to thermal conductivity tensor is expressed as~\cite{Simoncelli_NP19,Isaeva_NC19}
\begin{equation}
\kappa_{\alpha\beta}^{\mathrm{nd}} = \frac{1}{V} \sum_{\lambda\lambda'} v_{\lambda\lambda'}^{\alpha}v_{\lambda\lambda'}^{\beta} \frac{c_\lambda + c_{\lambda'}}{2} \Gamma_{\lambda\lambda'}
\end{equation}
where \(v_{\lambda\lambda'}^{\alpha}\) is the off-diagonal group velocity. The off-diagonal scattering rate \(\Gamma_{\lambda\lambda'}\) is defined as
\begin{equation}
\Gamma_{\lambda\lambda'} = \frac{\Gamma_\lambda + \Gamma_{\lambda'}}{(\omega_\lambda - \omega_{\lambda'})^2 + (\Gamma_\lambda + \Gamma_{\lambda'})^2}
\end{equation}
where \(\Gamma_\lambda\) is the scattering rate of phonon mode \(\lambda\) and the detailed expression for different scattereing processes can be found in Ref.~\onlinecite{Batista_PRB25}.


The key requirement to evaluate both the diagonal and off-diagonal contributions is the harmonic and anharmonic interatomic force constants (IFCs). To obtain these, we employ TDEP with efficient stochastic phase-space sampling scheme~\cite{Shulumba_PRB17}. This method iteratively refines the force constants by approximating atomic displacement distributions in the canonical ensemble and improving the approximation with actual forces. These forces are calculated using density functional theory (DFT) as implemented in the VASP package~\cite{VASP196,VASP299}. The exchange-correlation functional used for the final results is PBESol~\cite{PBEsol}, with additional comparisons made using the LDA~\cite{LDA} and PBE~\cite{PBE} functionals.

All DFT calculations were performed using a plane-wave energy cutoff of 500 eV and $\Gamma$-point sampling for the Brillouin zone. The convergence criteria for the electronic self-consistent cycle was set to $10^{-8}$ eV. A $5\times 5\times 5$ supercell containing 250 atoms was used to construct the second-, third-, and fourth-order IFCs. Cutoff radii of 6~\AA\ and 4~\AA\ were used for the third- and fourth-order IFCs, respectively. A detailed convergence analysis of the cutoff values for IFCs and the grid density used in the thermal conductivity calculations is provided in the Supplemental Material~\cite{supple}. For the pressure dependence study, the same workflow as described above was followed, except that forces were calculated using VASP's on-the-fly machine learning force field (MLFF)~\cite{VASP-MLFF}. Details of the MLFF generation are provided in the Supplemental Material~\cite{supple}.

\section{Results and discussions}

We begin by analyzing the phonon dispersion across various cases. Next, we examine the lattice thermal conductivity ($\kappa_{l}$), emphasizing the importance of temperature-dependent IFCs and higher-order phonon-phonon interactions. Finally, we investigate the unusual pressure dependence of $\kappa_{l}$.

\subsection{Lattice dynamics}

\begin{figure}[t]
     \centering
     \includegraphics[width=0.98\linewidth]{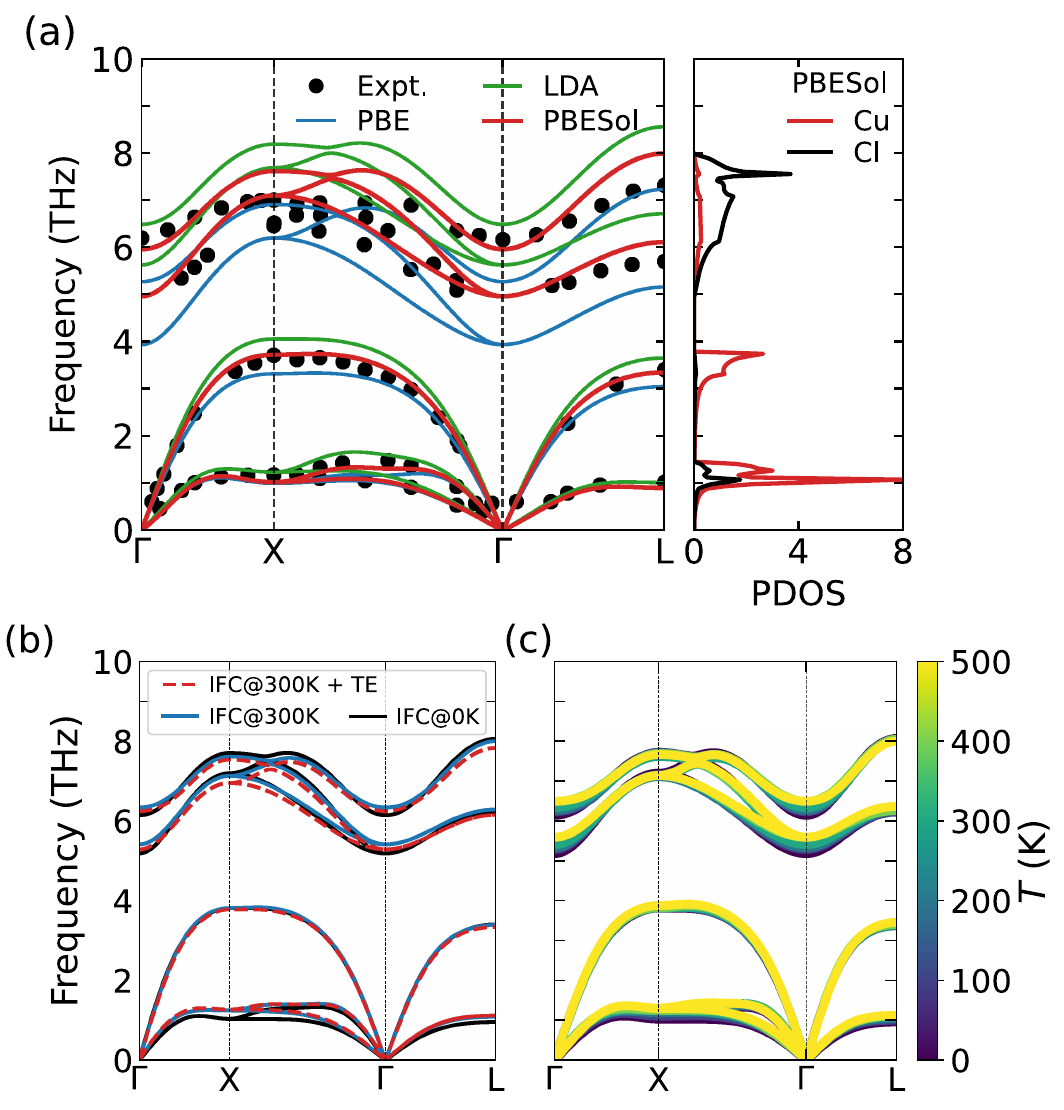}
     \caption{Calculated phonon dispersion and phonon density of states (PDOS) of \ce{CuCl}: (a) Phonon dispersion at \SI{0}{\kelvin} calculated using different exchange-correlation functionals, compared with experimental data measured at \SI{4.2}{\kelvin} (Ref.~\onlinecite{Prevot_JPCSSP77}). (b) Phonon dispersion at \SI{300}{\kelvin} without and with thermal expansion (TE), showing its minimal effect. The dispersion obtained using IFCs at 0 K is also shown for direct comparison. (c) Phonon dispersion at different temperatures, showing mode hardening with increasing temperature. The results in panels (b) and (c) were obtained using the PBEsol functional.}
     	\label{ph1}
 \end{figure}

$\kappa_{l}$ strongly depends on the features of phonon dispersion. We first compare the phonon dispersion calculated at \SI{0}{\kelvin} using various exchange-correlation functionals with experimental data measured at \SI{4.2}{\kelvin}~\cite{Prevot_JPCSSP77}, shown in Fig.~\ref{ph1}. The corresponding lattice parameters are summarized in the Supplemental Material~\cite{supple}. PBE shows the best agreement with experiment for the lattice parameter ($a =\SI{5.43}{\AA}$ vs. experimental $a =\SI{5.424}{\AA}$), while PBEsol ($a =\SI{5.30}{\AA}$) and LDA ($a =\SI{5.22}{\AA}$) underestimate it by approximately 2.2\% and 3.7\%, respectively. However, among the functionals, PBEsol provides reasonably good agreement with the phonon dispersion, particularly in the acoustic region, which typically dominates the heat transport. Based on this agreement, we choose PBEsol for all subsequent calculations and later confirmed its reliability by comparing $\kappa_{l}$ across different functionals. Since \ce{CuCl} is highly anharmonic~\cite{Hanson_PRB74}, a significant temperature effect on its phonon dispersion is expected, and we observe that the phonon modes hardens with increasing temperature, specifically near the high-symmetry X and $\Gamma$ points [Fig.~\ref{ph1} (c)]. Furthermore, lattice thermal expansion at \SI{300}{\kelvin} has a minor effect on the phonon dispersion. Overall, these results highlight the importance of incorporating temperature-dependent interatomic force constants (IFCs) for accurate predictions of $\kappa_{\mathrm{l}}$. Another notable observation is the flat nature of the acoustic modes around the high-symmetry X and L points, as reflected by the localized phonon density of states. This flatness suggests significant four-phonon scattering and its subsequent impact on $\kappa_{\mathrm{l}}$~\cite{Xie_PRL20}.

\begin{figure}[t]
     \centering
     \includegraphics[width=1\linewidth]{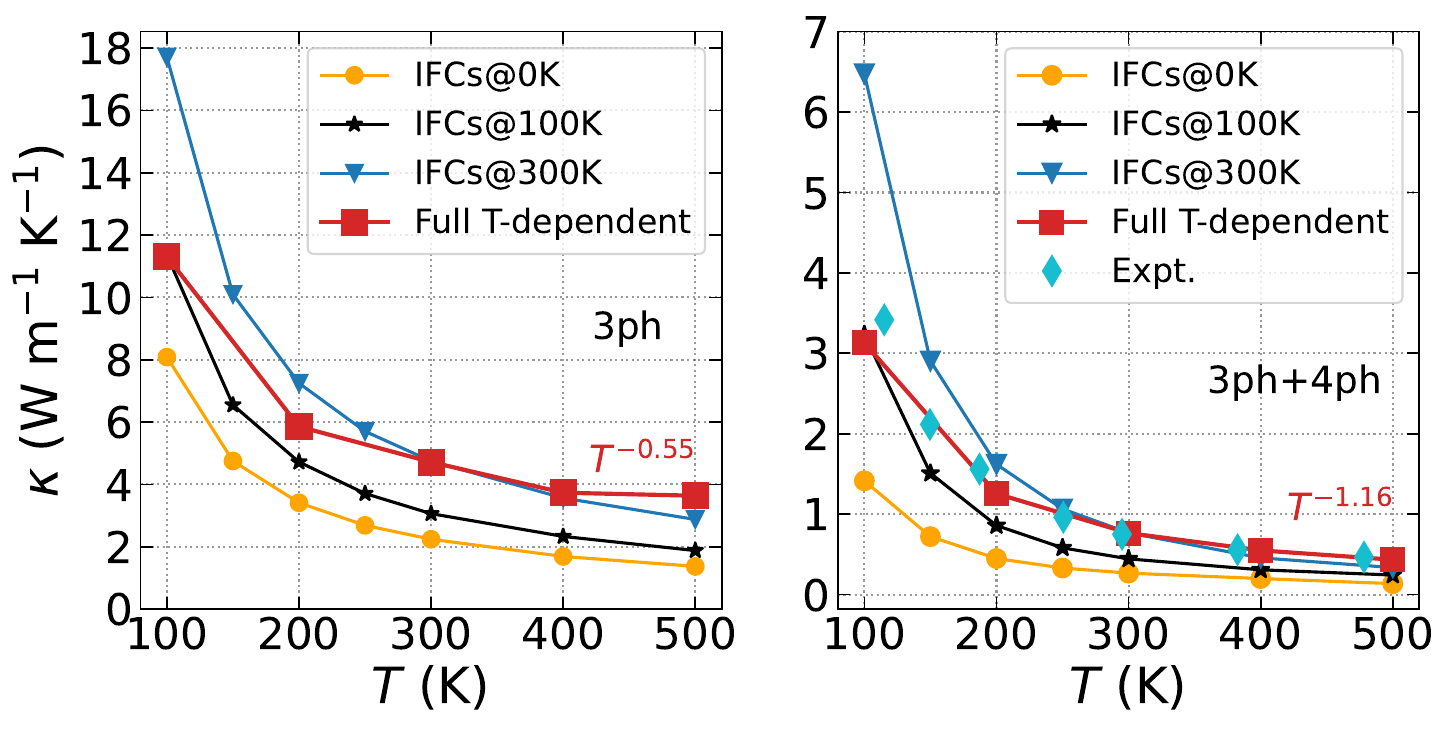}
     \caption{Calculated lattice thermal conductivity of \ce{CuCl} as a function of temperature using IFCs obtained at selected temperatures and fully temperature-dependent IFCs. Results are presented for (a) three-phonon (3ph) interactions only, and (b) combined three- and four-phonon (3ph + 4ph) interactions. Experimental $\kappa_{l}$ data are taken from Slack et al.~\cite{Slack_PRB82}, where the ambient-pressure values are obtained by extrapolating measurements in the range 0.5-2.8 GPa.}
     \label{kappa}
\end{figure}


\subsection{Impact of temperature-dependent IFCs on $\kappa_{l}$}

\begin{figure}[t]
     \centering
     \includegraphics[width=0.95\linewidth]{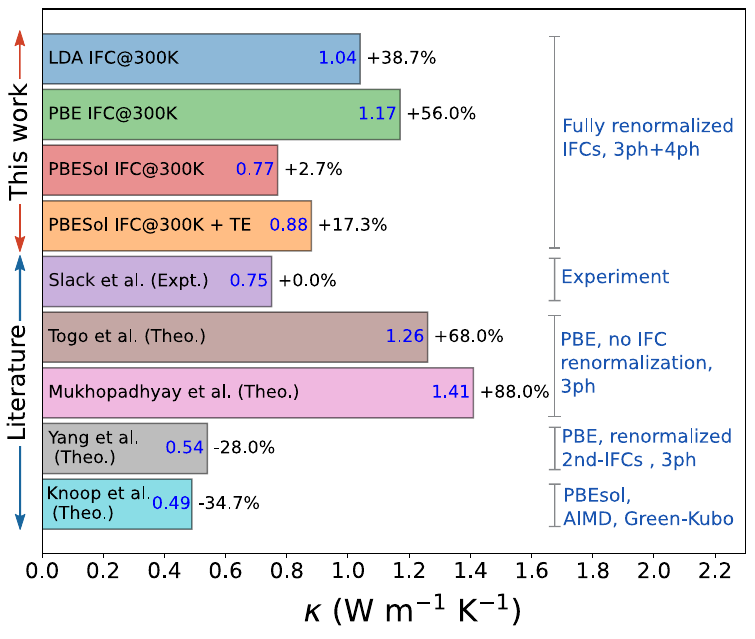}
     \caption{Comparison of \ce{CuCl} lattice thermal conductivity for different exchange-correlation functionals and computational methods. In all our cases, IFC renormalization and four-phonon scattering are included. `TE' denotes the correction for lattice thermal expansion. Literature data are also presented, including experimental results~\cite{Slack_PRB82}, BTE calculations without IFC renormalization and four-phonon scattering~\cite{Togo_PRB15,Mukhopadhyay_PRB17}, calculations with IFC renormalization on second-order terms only (without four-phonon scattering)~\cite{Yang_SCPMA23}, and MD simulations~\cite{Knoop_PRL23}.}
     \label{kappa-bar}
 \end{figure}

We analyze $\kappa_{l}$ using IFCs obtained at different temperatures. Figure~\ref{kappa} presents the calculated $\kappa_{l}$ considering only three-phonon scattering and both three- and four-phonon scattering. At \SI{300}{\kelvin}, $\kappa_{l}$ increases when IFCs at \SI{0}{\kelvin} are substituted with those at \SI{300}{\kelvin}, for both the scattering cases. A similar increase is also observed at other temperatures, consistent with trends reported in other materials~\cite{Klarbring_PRL20,Li_PRB23}. Using \SI{0}{\kelvin} IFCs, $\kappa_{l}$ at \SI{300}{\kelvin} decreases from \SI{2.24} to \SI{0.27}{\watt\per\meter\per\kelvin} upon inclusion of four-phonon scattering. Similarly, using \SI{300}{\kelvin} IFCs, it decreases from \SI{4.75} to \SI{0.77}{\watt\per\meter\per\kelvin}. This corresponds to approximately 88\% and 84\% reductions in $\kappa_{l}$ with the inclusion of four-phonon scattering when using \SI{0}{\kelvin} and \SI{300}{\kelvin} IFCs, respectively. Thus, regardless of whether temperature-dependent IFCs are used, $\kappa_{l}$ exhibits a significant decrease compared to other binary compounds with the zincblende structure~\cite{Ravichandran_PRX20,Xia_PRX20}. Note that the coherent contribution to $\kappa_{\mathrm{l}}$ is negligible (approximately \SI{0.04}{\watt\per\meter\per\kelvin} for three-phonon and \SI{0.06}{\watt\per\meter\per\kelvin} for combined three- and four-phonon scattering). This behavior arises from the simple crystal structure of CuCl, which has only a few well-separated phonon branches. As a result, the large phonon interband spacing across the Brillouin zone limits interband coupling, unlike in materials with many atoms per unit cell and complex phonon dispersion, where strong interband mixing leads to notable coherent contributions~\cite{Simoncelli_NP19,Simoncelli_PRX22}. Overall, the calculated $\kappa_{\mathrm{l}}$ shows excellent agreement with experimental values across all temperatures [Fig.~\ref{kappa}].

To quantify the effect of IFC renormalization, we calculate $\kappa_l$ using various combinations of second-, third-, and fourth-order IFCs obtained at \SI{0}{\kelvin} and \SI{300}{\kelvin}. The results reveal that temperature-induced renormalization affects all orders of IFCs and is crucial for the accurate estimation of $\kappa_l$, with particularly pronounced effects observed for the second- and fourth-order interactions (see Supplemental Material~\cite{supple}).

Another important aspect is the temperature dependence of $\kappa_{l}$. When only three-phonon scattering is considered, $\kappa_{l}$ follows a weak temperature dependence, approximately $\sim T^{-0.55}$ [Fig.~\ref{kappa} (a)], deviating from the conventional $\sim T^{-1}$ behavior. Inclusion of four-phonon scattering strengthens temperature dependence to about $\sim T^{-1.16}$ [Fig.~\ref{kappa} (b)], closer to the typical $\sim T^{-1}$ trend observed in most materials. This change upon inclusion of four-phonon scattering is consistent with earlier reports on other low-$\kappa_{l}$ systems~\cite{Yue_npjCM23,Wang_APL24}, also emphasizing the essential role of four-phonon interactions in lattice thermal transport.

We employed the PBEsol functional for calculating $\kappa_{l}$​, based on its better agreement with experimental phonon dispersion compared to PBE and LDA, which exhibit considerable mismatches. This choice is supported by the computed $\kappa_{l}$ discussed earlier. However, for further validation, we calculate $\kappa_{l}$ using different exchange-correlation functionals and compared it with previously reported theoretical and experimental results, as displayed in Fig.~\ref{kappa-bar}. Both PBE and LDA significantly overestimate $\kappa_{l}$​. This observation is consistent with earlier reports on other materials, where PBEsol shows smaller deviations from experiment than other functionals~\cite{Wei_PRB24}. Furthermore, incorporating thermal expansion results in an approximately 17\% higher $\kappa_{l}$ than the experimental value [Fig.~\ref{kappa-bar}]. Overall, the good agreement between our results and experiment, in contrast to significant discrepancies in earlier theoretical predictions that neglected phonon renormalization and four-phonon scattering, highlights the essential role of temperature-dependent IFCs and higher-order phonon interactions in accurately modeling lattice thermal conductivity.

\begin{table}[b]
    \centering
    \renewcommand{\arraystretch}{1.2} 
    \setlength{\tabcolsep}{5pt} 
    \footnotesize 
    \begin{tabular}{|c|c|c|c|c|}
    \hline
    \hline
    System & $\kappa_{\mathrm{3ph}}$ & $\kappa_{\mathrm{3ph+4ph}}$ & $\kappa_{\mathrm{3ph+4ph}}^{TE}$ & $\kappa_{\mathrm{expt.}}$ \\ \hline
    
    \ce{CuCl}  & 4.75  & 0.76  & 0.88  & 0.75~\cite{Slack_PRB82} 0.84~\cite{Haynes_CRC14} \\
    \ce{CuBr}  & 7.38  & 1.35  & 1.35  & 1.25~\cite{Haynes_CRC14}\\
    \ce{CuI}   & 10.92  & 3.94  & 3.55  & 1.68~\cite{Haynes_CRC14} \\

    \hline
    \hline
    \end{tabular}
    \caption{Calculated lattice thermal conductivity $\kappa_{\mathrm{l}}$ at \SI{300}{\kelvin}. `TE' denotes the case where lattice thermal expansion is taken into account.}
    \label{tab1}
\end{table}

 \begin{figure}[t]
     \centering
     \includegraphics[width=0.90\linewidth]{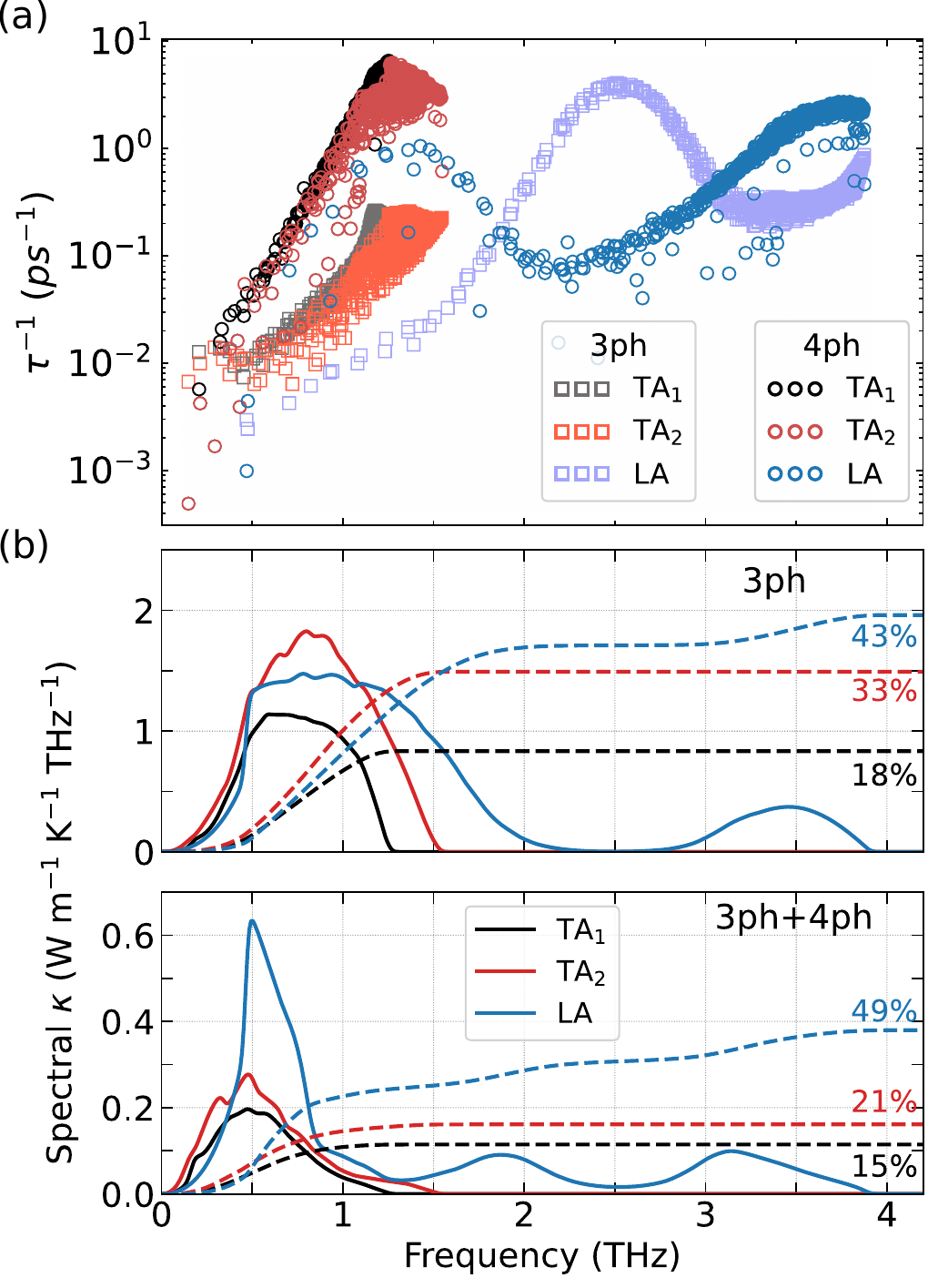}
     \caption{(a) Calculated three-phonon (3ph) and four-phonon (4ph) scattering rates for acoustic modes at 300 K. (b) Spectral thermal conductivity and its cumulative sum, with percentage contributions of each mode to the total thermal conductivity.}
     \label{sr}
 \end{figure}

To further validate the reliability of our approach, we computed $\kappa_{l}$ for \ce{CuBr} and \ce{CuI}, with results presented in Table~\ref{tab1}. For \ce{CuBr}, the calculated $\kappa_{l}$ agrees well with experimental value, and the effect of thermal expansion is found to be minimal. In the case of \ce{CuI}, however, our predicted $\kappa_{l}$ is somewhat higher than the reported experimental value, likely due to defects present in the measured samples, as noted previously~\cite{Knoop_PRL23}. These findings further emphasize that the excellent agreement of our calculated $\kappa_{l}$ results from a thorough and accurate assessment, confirming the robustness of our approach for low-$\kappa_{l}$ systems.

To elucidate the microscopic mechanisms underlying the ultralow $\kappa_l$ of \ce{CuCl}, we analyze mode-resolved three- and four-phonon scattering rates, along with the mode-resolved spectral contributions to $\kappa_l$, as shown in Fig.~\ref{sr}. Notably, the four-phonon scattering rates for the transverse acoustic (TA) modes are nearly a few orders of magnitude higher than the corresponding three-phonon scattering rates. This pronounced scattering by TA phonons occurs because of the large phonon population and the flat, nondispersive nature of the TA and longitudinal acoustic (LA) phonon branches near the high-symmetry X and L points, which favors resonance in four-phonon scattering processes~\cite{Kanellis_PRL86,Xie_PRL20}. From the mode-resolved spectral contributions, we find that low-frequency phonons below \SI{1.5}{\tera\hertz} dominate $\kappa_l$. The contribution of TA modes to the total $\kappa_l$ is approximately 51\% when only three-phonon scattering is included, but it decreases to 36\% when both three-phonon and four-phonon processes are considered.
Since phonons in this frequency range dominate the heat transport, the strong four-phonon scattering of TA phonons plays a significant role in reducing $\kappa_l$.

Another notable observation is the significant contribution of LA modes to $\kappa_l$, accounting for nearly 50\%, with approximately 20\% of the contribution arising from modes above \SI{1.5}{\tera\hertz}. This stems from competing scattering between three-phonon and four-phonon processes, which reduce the overall scattering rates near their crossover frequency region just below \SI{2}{\tera\hertz} and above \SI{3}{\tera\hertz}. Consequently, pronounced LA contributions to $\kappa_l$ emerge in these frequency ranges. This competition arises from energy and momentum conservation constraints associated with different scattering processes~\cite{Ravichandran_PRX20}. For instance, the peak around \SI{2.5}{\tera\hertz} in the three-phonon LA scattering originates from the sizable gap between TA and LA modes, which forbids many three-phonon processes and restricts three-phonon processes to a narrow frequency range ($2-$\SI{3}{\tera\hertz}) (see Supplemental Material~\cite{supple}). Similarly, four-phonon processes become dominant at higher frequencies, producing a peak in the LA four-phonon scattering rate between $3-$\SI{4}{\tera\hertz}. Thus, strong LA phonon scattering also plays a role in suppressing $\kappa_l$.

\subsection{Pressure dependence of $\kappa_{l}$}

\begin{figure}[t]
     \centering
     \includegraphics[width=0.98\linewidth]{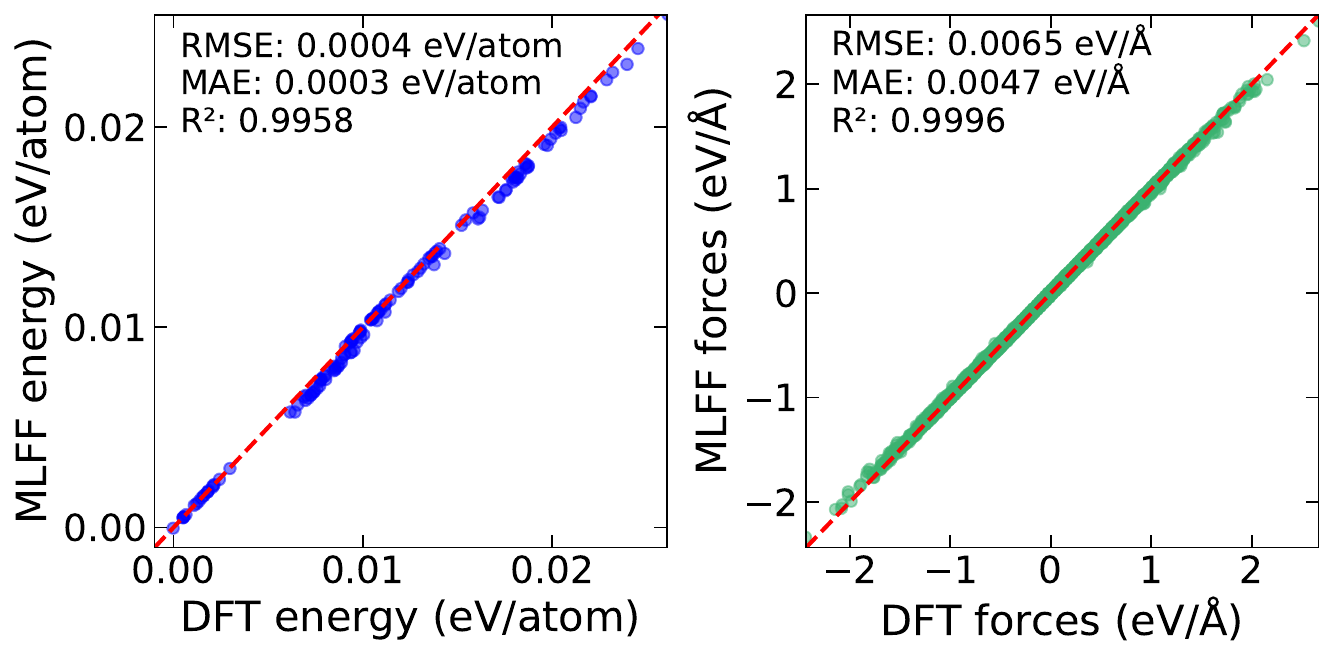}
     \caption{Comparison of total energies and atomic forces predicted by DFT and MLFF. The accuracy of the MLFF is quantified using RMSE, MAE, and $R^2$ metrics, as given in the plot.}
     \label{dft-mlff}
 \end{figure}

\begin{figure}[t]
     \centering
     \includegraphics[width=1.0\linewidth]{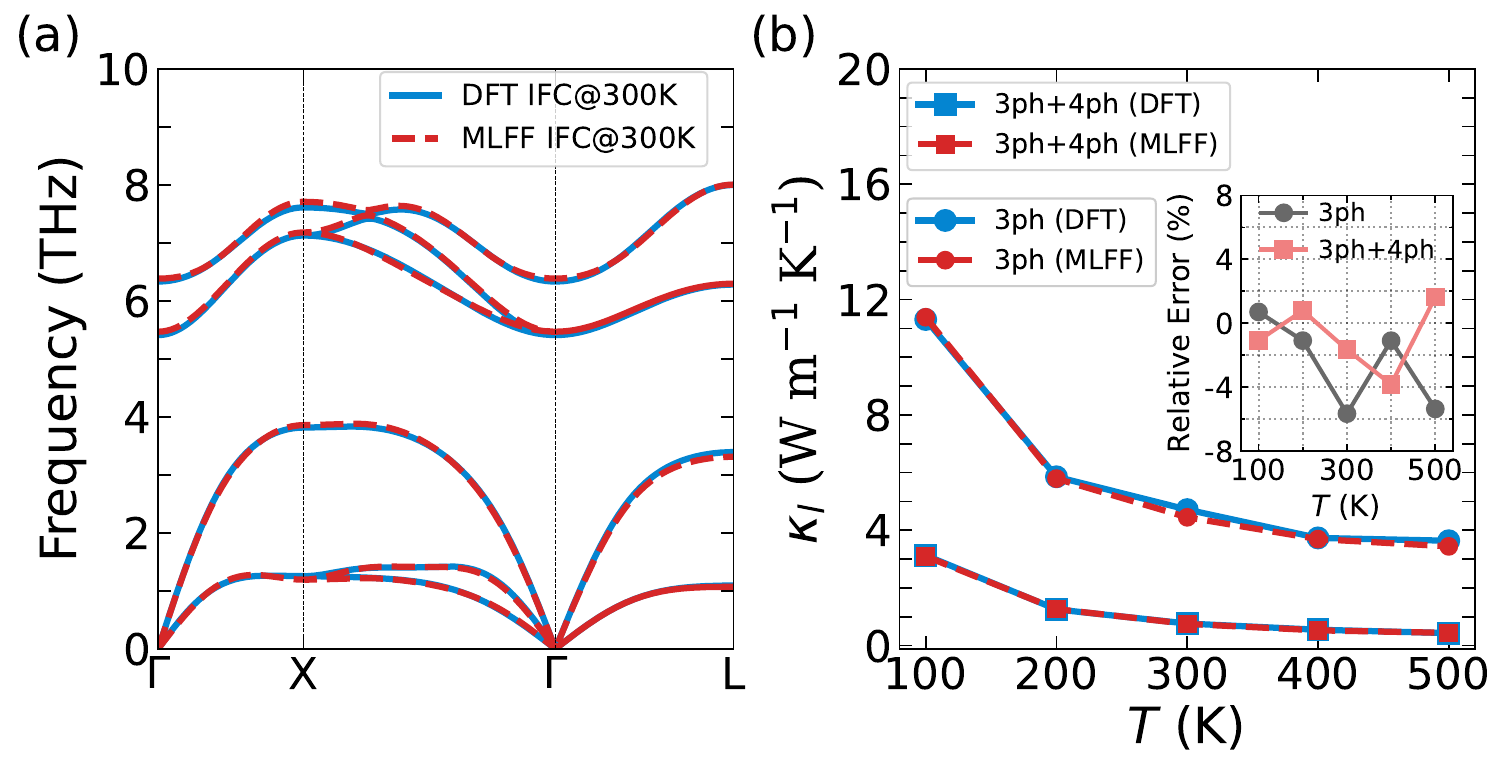}
     \caption{(a) Comparison of the phonon dispersion of \ce{CuCl} at 300 K obtained from DFT and MLFF. (b) Temperature-dependent lattice thermal conductivity calculated using IFCs from DFT and MLFF, considering both three-phonon (3ph) and four-phonon (3ph+4ph) interactions. The inset shows the relative error ($\frac{\kappa^{DFT} - \kappa^{MLFF}}{\kappa^{DFT}} \times 100\%$) in thermal conductivity between DFT and MLFF results.}
     \label{ph2}
 \end{figure}

In this section, we investigate the unusual pressure dependence of $\kappa_{l}$ for \ce{CuCl}. Computing $\kappa_{l}$ requires both harmonic and anharmonic IFCs, which are typically obtained from numerous first-principles DFT calculations. To overcome this challenge and substantially reduce the computational cost, we employ a machine learning force field (MLFF) approach. In the following, we provide a detailed validation of the MLFF prior to calculating $\kappa_{l}$ under various pressures.

Figure~\ref{dft-mlff} compares the energies and forces computed using DFT and those calculated using the MLFF on a test set. This test set comprises structures sampled across various temperatures, generated by emulating a canonical ensemble using TDEP under different pressures. The low root-mean-square errors (RMSEs) in both energies and forces demonstrate the accuracy and reliability of the MLFF. Furthermore, Fig.~\ref{ph2} (a) shows that the phonon dispersion calculated using MLFF closely reproduces the DFT results, confirming the capability of MLFF to accurately describe atomic motion in \ce{CuCl}. Although energy and force comparisons are commonly used to validate MLFF, we extend our validation by comparing $\kappa_{l}$ at ambient pressure, calculated using IFCs obtained from both DFT and MLFF, as illustrated in Fig.~\ref{ph2} (b). The deviation in $\kappa_{l}$ between the two methods remains within 6\% across different temperatures, regardless of the order of scattering processes considered.
This close agreement in $\kappa_{l}$ values demonstrates the robustness and transferability of the MLFF, confirming its reliability for accurate estimation of $\kappa_{l}$ under pressure.

\begin{figure}[t]
     \centering
     \includegraphics[width=0.9\linewidth]{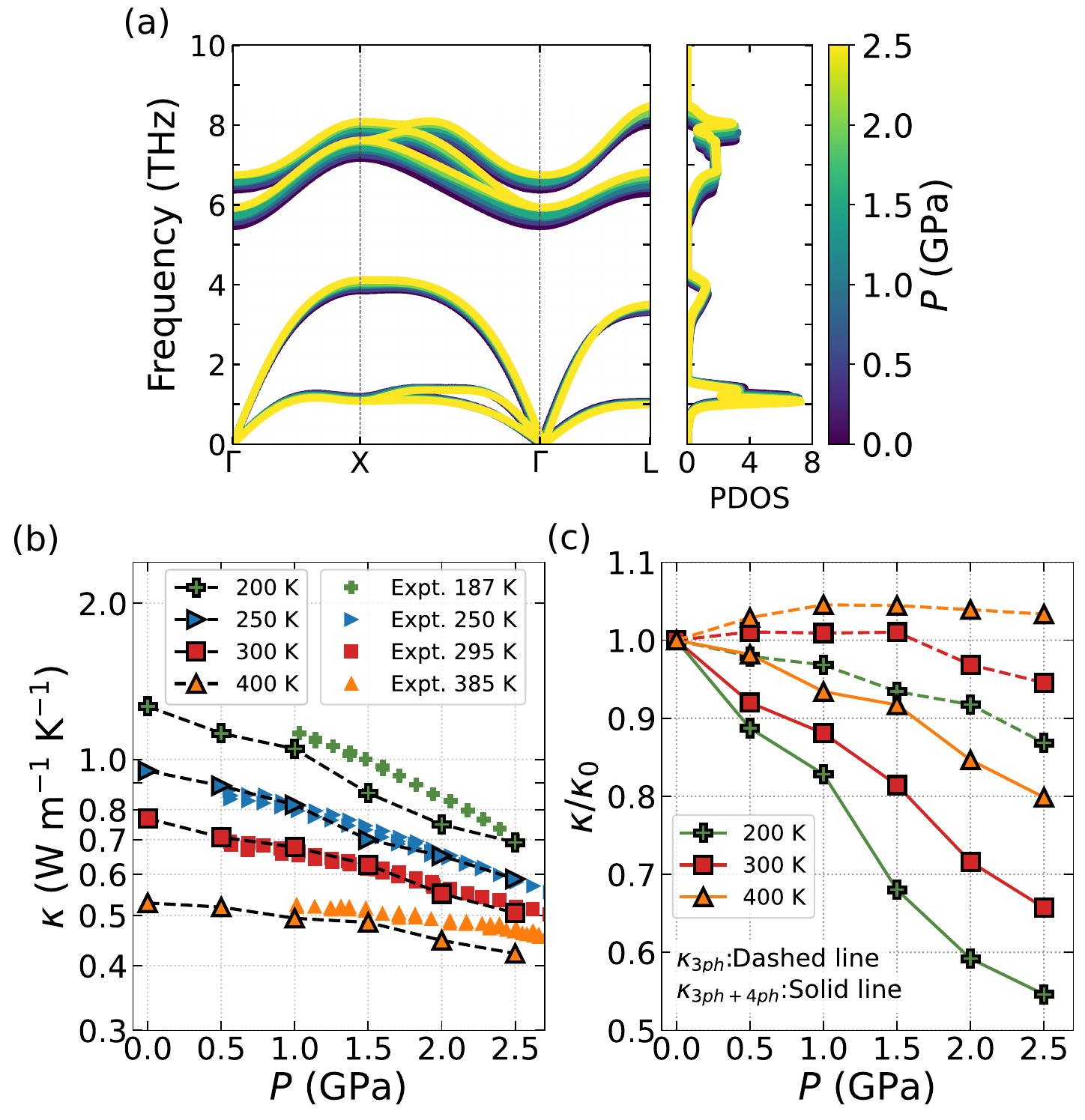}
     \caption{(a) Calculated phonon dispersion and phonon density of states (PDOS) of \ce{CuCl} at 300 K under different pressures. (b) Calculated lattice thermal conductivity as a function of pressure at different temperatures, shown in comparison with experimental results from Ref.~\onlinecite{Slack_PRB82}. (c) Scaled lattice thermal conductivity relative to the 0 GPa value, considering both three-phonon (3ph) and four-phonon (4ph) scattering.}
     \label{kappa-p}
 \end{figure}

Figure~\ref{kappa-p} (a) shows the phonon dispersion of \ce{CuCl} under applied pressure. Typically, pressure leads to phonon mode hardening due to shortened bond lengths. However, in \ce{CuCl}, a softening of the transverse acoustic (TA) modes at the X and L points is observed. At the same time, the remaining phonon modes exhibit the expected hardening behavior under pressure, as also reflected in the phonon density of states. This softening behavior of TA modes arises from the unusual nature of the out-of-phase vibrations at those points (see Supplemental Material~\cite{supple}). Under pressure, these vibrations induce shear distortions rather than directly compressing the bonds, indicating weaker interatomic interactions. A similar softening behavior under pressure has been reported in other binary compounds with the zincblende structure, such as BAs~\cite{Ravichandran_NC19}, BP~\cite{Ravichandran_NC21}, and HgTe~\cite{Ouyang_15}. However, the softening occurs at much higher applied pressures in those materials. Moreover, the latter compounds are characterized by relatively strong bonding, leading to moderate to very high thermal conductivity, unlike \ce{CuCl}, which exhibits an ultralow thermal conductivity. Interestingly, the pressure dependence of thermal conductivity in these materials shows anomalous trends, different from the traditional belief that thermal conductivity increases with pressure. In contrast, \ce{CuCl} is highly anharmonic, and the pressure-induced softening of its TA mode appears to play a significant role in its $\kappa_l$, as experimental observations show a continuous decrease in $\kappa_l$ with increasing pressure~\cite{Slack_PRB82}.

The prediction of pressure dependence of $\kappa_l$ can vary significantly depending on the order of phonon scattering processes included in the BTE calculation~\cite{Ravichandran_NC19,Ravichandran_NC21}. For example, in the case of BAs~\cite{Ravichandran_NC19}, previous studies have shown that including four-phonon scattering significantly alters the pressure-dependent trend of $\kappa_l$. Instead of a continuous decrease, the trend becomes anomalous, where $\kappa_l$ initially increases at low pressures and then decreases after reaching a critical pressure. This anomaly is further supported by experimental measurements~\cite{Li_N22}, highlighting the importance of considering four-phonon scattering. However, a few recent measurements have reported unexpectedly large values of $\kappa_{l}$ for BAs at ambient pressure, indicating the need for further detailed investigations under both ambient and high-pressure conditions.~\cite{Hou_PRB25,Ange_MT25}.

Given this, we have computed $\kappa_l$ by including all relevant and plausible phonon-phonon interactions to ensure an accurate estimation. Figure~\ref{kappa-p} (b) presents our calculated $\kappa_l$ as a function of pressure at various temperatures. Our results show good agreement with previously reported experimental data. To highlight the importance of four-phonon scattering, we present the normalized $\kappa_{l}$ considering only three-phonon scattering and both three-, and four-phonon scattering in Figure~\ref{kappa-p} (c). As expected, using only three-phonon scattering yields a trend that deviates from experimental observations, while the inclusion of four-phonon scattering captures the proper pressure-dependent behavior. Notably, $\kappa_{l}$ exhibits a stronger pressure dependence at lower temperatures, which gradually weakens as the temperature increases. This behavior can be attributed to the faster weakening of four-phonon scattering under pressure~\cite{Kundu_PRL24}. It is driven by the intrinsically larger weighted phase space of four-phonon processes, which are more sensitive to $\kappa_{l}$ in \ce{CuCl} than three-phonon processes. Note that thermal expansion is not included in the calculations of $\kappa_{l}$ under pressure, as its effect is found to be small at ambient pressure and becomes negligible with increasing pressure (see Supplemental Material~\cite{supple}).




\begin{figure}[t]
\centering
\includegraphics[width=0.98\linewidth]{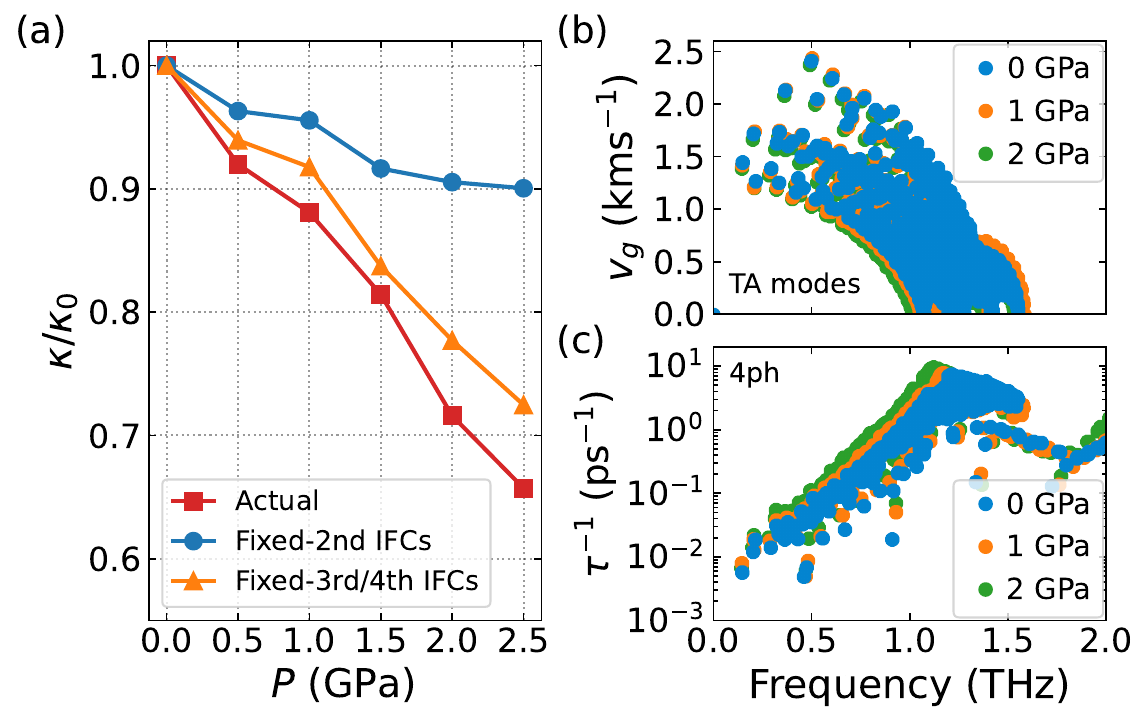}
\caption{(a) Scaled lattice thermal conductivity at \SI{300}{K} as a function of pressure, normalized to its value at \SI{0}{GPa}. `Fixed-2nd IFCs' refers to calculations where harmonic IFCs are fixed at \SI{0}{GPa}, while anharmonic IFCs vary with pressure. Conversely, `Fixed-3rd/4th IFCs' denotes calculations where anharmonic IFCs are fixed at \SI{0}{GPa}, and harmonic IFCs vary with pressure. (b) Group velocity ($v_g$) and (c) four-phonon scattering rates of TA modes at \SI{300}{K}, both of which increase under pressure.}
\label{kappa-pp}
\end{figure}

To further investigate the underlying mechanisms, we analyze harmonic and anharmonic quantities to identify the factors driving the continuous decrease of $\kappa_{l}$ under pressure. An increase in the harmonic contribution, proportional to the product of the specific heat capacity $C_v$​ and the square of the phonon group velocity $v_\text{g}^2$​, enhances $\kappa_{l}$​, while an increase in anharmonicity reduces it. Under pressure, the group velocities $v_{g}$​ of the LA and optical modes increase, whereas those of the low-frequency TA modes decrease (see Supplemental Material~\cite{supple}), resulting in a mixed effect on $\kappa_{l}$​​. Similarly, the Grüneisen parameter, which reflects anharmonicity, decreases significantly for the TA modes while remaining nearly constant for the LA mode (see Supplemental Material~\cite{supple}). Furthermore, the four-phonon scattering rates of TA modes increase under pressure (see Supplemental Material~\cite{supple}), while those of LA modes show mixed behavior, with both increases and decreases at different frequencies. From the mode-projected $\kappa_{l}$​ at \SI{0}{GPa}, it has been seen that the LA modes contribute nearly 50\%, which remains largely unchanged under pressure. Therefore, the decrease in $v_{g}$​ and increase in four-phonon scattering for TA modes alone cannot fully explain the reduction in $\kappa_{l}$. To elucidate further, we analyze the mode-projected $\kappa_{l}$ under pressure, which reveals that four-phonon scattering rates for both TA and LA modes increase significantly in the frequency range most relevant to $\kappa_{l}$. Overall, the effect of increasing scattering rates, i.e., anharmonicity, is evident. However, the group velocity of TA modes decreases while it increases for LA modes.

\begin{figure}[t]
     \centering
     \includegraphics[width=0.90\linewidth]{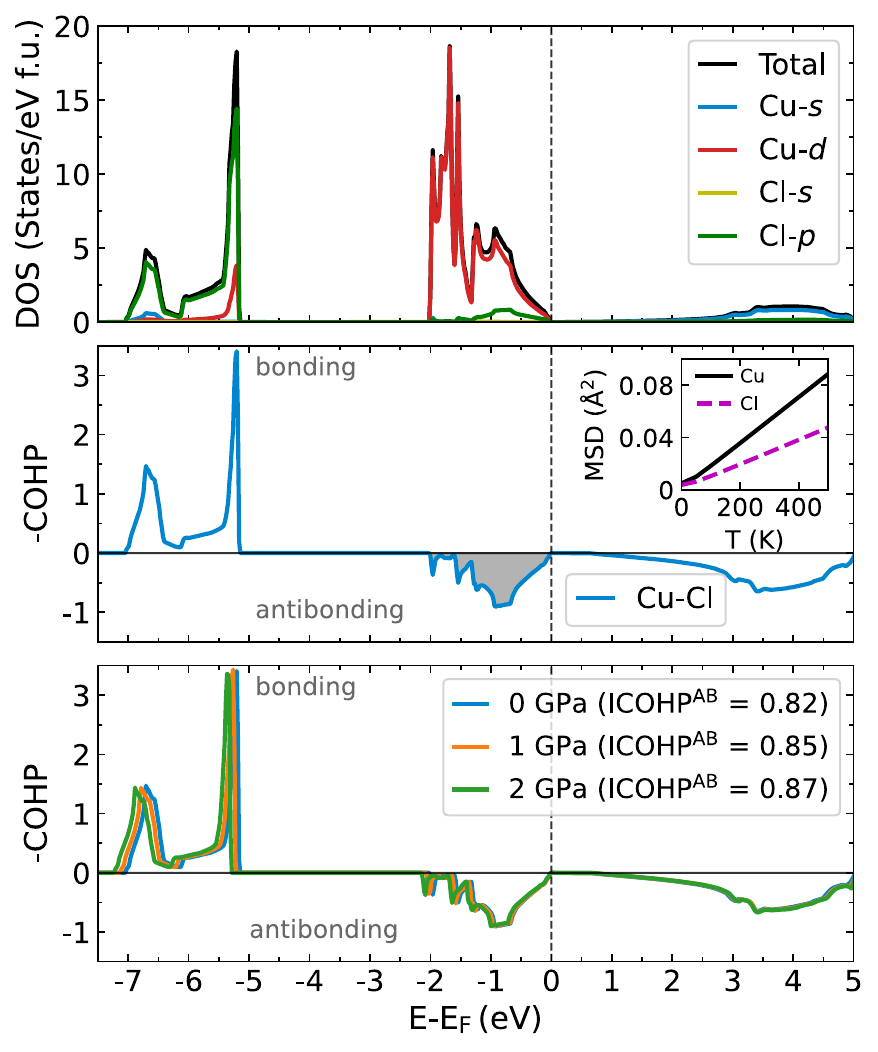}
     \caption{Calculated electronic density of states (DOS) and crystal orbital Hamilton population (COHP) of \ce{CuCl} at ambient and applied pressures. E$_\mathrm{F}$ denotes the Fermi energy. (inset) Calculated mean square displacements (MSD) of Cu and Cl along the bond direction at \SI{0}{GPa}. The integrated COHP for antibonding occupied states (ICOHP$^{\mathrm{AB}}$) increases with pressure.}
     \label{dos}
 \end{figure}

To quantify the impact of harmonic and anharmonic contributions under pressure, we perform cross calculations of $\kappa_{l}$ by separately fixing harmonic (2nd-order) and anharmonic (3rd- and 4th-order) IFCs across pressures, as shown in Fig.~\ref{kappa-pp} (a). When the 3rd- and 4th-order IFCs are fixed at \SI{0}{GPa}, $\kappa_{l}$ decreases more slowly compared to the case where the 2nd-order IFCs are fixed at \SI{0}{GPa}. This implies that changes in the harmonic contribution have a greater impact on $\kappa_{l}$ under pressure than changes in the anharmonic contribution, although both are essential for capturing the actual trend. The reduction in group velocity and the increase in four-phonon scattering in the low-frequency regime [Figure~\ref{kappa-pp} (b) and (c)] primarily drive the trend.  These observations further emphasize the importance of phonon renormalization and inclusion of four-phonon  interactions in accurately determining the pressure dependence of $\kappa_{l}$.

\subsection{Electronic DOS and COHP}

So far, we have explored the factors that directly influence the estimation of $\kappa_{l}$ and explained the origins of its ultralow value as well as its unusual pressure dependence. Among these factors, the interatomic force constants are fundamental parameters, as they serve as the primary inputs for thermal conductivity calculations and reflect the strength and nature of atomic bonding. Weak bonding typically leads to low thermal conductivity. Thus, to gain insight into the bonding characteristics, we present the electronic density of states (DOS) and crystal orbital Hamilton population (COHP)~\cite{LOBSTER}, shown in Fig~\ref{dos}. The Cu-$d$ orbitals largely contribute to the top of the valence band, while the Cl-$p$ orbitals dominate the lower energy region. A weak overlap between Cu-$d$ and Cl-$p$ states is observed just below the Fermi energy. This weak hybridization typically results in the formation of bonding and antibonding states, consistent with observations in other materials~\cite{Das_JACS23,Yuan_JACS23}. However, in the case of \ce{CuCl}, the weak $p$-$d$ hybridization results predominantly in antibonding states. These antibonding states located just below the Fermi level indicate a dominant ionic character, resulting in overall weaker bonding. Also, the large mean square displacements (MSD) of Cu atoms further confirm the weak bonding characteristics in the system. This observation aligns with trends seen in other materials exhibiting ultralow $\kappa_{l}$. Under pressure, the antibonding states slightly extend to lower energies, and the calculated integrated value of antibonding states (ICOHP$^{\text{AB}}$) increases, which describes the weakening of bonding. This weakening of bonding can be linked to the softening of the transverse acoustic modes under pressure, which subsequently reduces $\kappa_{l}$.

\section{Conclusions}

In summary, we have investigated the $\kappa_{l}$ of \ce{CuCl} using a comprehensive framework that incorporates interatomic force constants (IFC) renormalization, four-phonon scattering, thermal expansion, and full iterative solutions of the Boltzmann transport equation. Our calculated $\kappa_{l}$ shows good agreement with reported experimental results, highlighting the significance of including four-phonon scattering and IFC renormalization to resolve discrepancies found in earlier studies. The effects of exchange-correlation functionals and thermal expansion on $\kappa_{l}$ are also highlighted. We have further investigated the unusual pressure dependence of $\kappa_{l}$. Our calculated $\kappa_{l}$ not only matches the experimental results at ambient pressure but also reproduces its trend under pressure at different temperatures. Microscopic analysis reveals that both the weakening of harmonic quantities (group velocity) and the increase in anharmonic quantities, notably the four-phonon scattering, reduces $\kappa_{l}$ under pressure. However, the weakening in harmonic quantities plays the dominant role, though both effects are essential for reproducing the experimentally observed trends under pressure. From a methodological viewpoint, we provide a clear workflow to compute $\kappa_{l}$ using MLFF combined with the TDEP. The reliability of the MLFF method has been thoroughly validated. Overall, our accurate calculations and their agreement with experimental results provide valuable insights into thermal transport in \ce{CuCl} and other strongly anharmonic materials.

\section*{ACKNOWLEDGMENTS}
A.K. thanks J. Klarbring and A. Castellano for helpful discussions during the project. A.K. and I.A. gratefully acknowledge the computational resources provided by the National Academic Infrastructure for Supercomputing in Sweden (NAISS) and the Swedish National Infrastructure for Computing (SNIC) at NSC. We acknowledge support from the Knut and Alice Wallenberg Foundation (Wallenberg Scholar grant no. KAW- 2023.0309), the Swedish Research Council (VR) Grant no. 2023-05358 and the Swedish Government Strategic Research Area in Materials Science on Functional Materials at Link¨oping University (Faculty Grant SFOMat-LiU No. 2009 00971). F.K. acknowledges support from the Swedish Research Council (VR) program 2020-04630 and the Swedish e-Science Research Centre (SeRC).

\section*{DATA AVAILABILITY}

The data supporting this work are available on Zenodo~\cite{data_cucl}. The dataset includes thermally displaced atomic positions, the corresponding forces, and the \texttt{ML\_AB} file, which contains training data generated from the VASP on-the-fly machine-learning force field.


%

\end{document}